\documentclass[conference]{IEEEtran}
\usepackage{hyperref}
\ifCLASSINFOpdf
   \usepackage[pdftex]{graphicx}
\else
\fi
\begin{document}
%
\title{An Overview of IEEE 802.15.6 Standard}

\author{\IEEEauthorblockN{Kyung Sup Kwak, Sana Ullah, and Niamat Ullah}
\IEEEauthorblockA{UWB-ITRC Center, Inha University\\
253 Yonghyun-dong, Nam-gu, Incheon (402-751), South Korea \\
Email: kskwak@inha.ac.kr, sanajcs@hotmail.com, niamatnaz@gmail.com}
}

%

\IEEEspecialpapernotice{(Invited Paper)}

\maketitle

\begin{abstract}
Wireless Body Area Networks (WBAN) has emerged as a key technology to provide real-time health monitoring of a patient and diagnose many life threatening diseases. WBAN operates in close vicinity to, on, or inside a human body and supports a variety of medical and non-medical applications. IEEE 802 has established a Task Group called IEEE 802.15.6 for the standardization of WBAN. The purpose of the group is to establish a communication standard optimized for low-power in-body/on-body nodes to serve a variety of medical and non-medical applications. This paper explains the most important features of the new IEEE 802.15.6 standard. The standard defines a Medium Access Control (MAC) layer supporting several Physical (PHY) layers. We briefly overview the PHY and MAC layers specifications together with the bandwidth efficiency of IEEE 802.15.6 standard. We also discuss the security paradigm of the standard.                       
\end{abstract}

%
\IEEEpeerreviewmaketitle

\section{Introduction}
\begin{figure}[!t]
\centering
\includegraphics[width=3.5in]{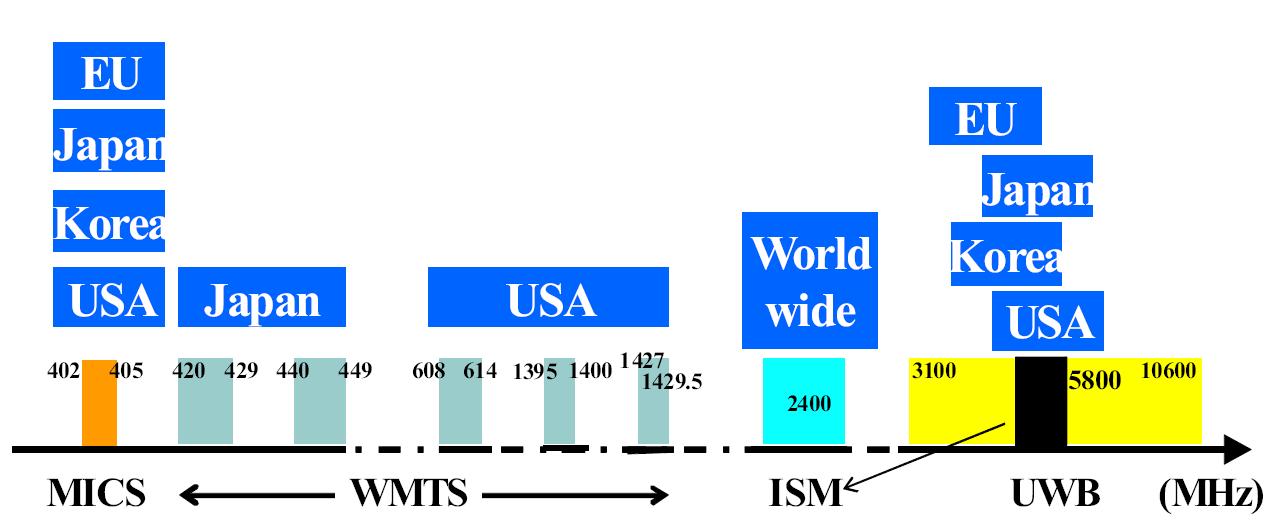}
\caption{Frequency bands for WBAN \cite{astrin}}
\label{fig:1}
\end{figure}

Wireless Body Area Network (WBAN) has great potential to revolutionize the future of healthcare technology. It has attracted a number of researchers both from the academia and industry in the past few years. WBAN supports a wide range of medical and Consumer Electronics (CE) applications. For example, WBAN provides remote health monitoring of a patient's state for a long period of time without any restriction on his/her normal activities \cite{sana}-\cite{chen}. For a successful implementation of WBAN, a standard model was required which would be able to address both medical and CE applications. IEEE 802 established a Task Group called IEEE 802.15.6\footnote{This paper is based on Draft.1 of the IEEE 802.15.6 standard (This paper was invited for presentation in ISABEL 2010 in Rome, Italy).} for the standardization of WBAN \cite{802.15.6}. Earlier, IEEE 802 had many success stories in developing international standards on wireless communication. Examples include IEEE 802.11 \cite{802.11}, IEEE 802.15.1 \cite{802.15.1} and IEEE 802.15.4 \cite{802.15.4} standards. The purpose of the IEEE 802.15.6 was to define new Physical (PHY) and Medium Access Control (MAC) layers for WBAN. The selection of the PHYs (frequency bands) were one of the most important issues. Generally, the available frequencies for WBANs are regulated by communication authorities in different countries. Fig. \ref{fig:1} shows a short summary of some of the frequency bands available for WBAN in different countries \cite{astrin}. Medical Implant Communications Service (MICS) band is a licensed band used for implant communication and has the same frequency range (402-405 MHz) in most of the countries. Wireless Medical Telemetry Services (WMTS) is a licensed band used for medical telemetry system. Both MICS and WMTS bandwidths do not support high data rate applications. The Industrial, Scientific and Medical (ISM) band supports high data rate applications and is available worldwide. However, there are high chances of interference as many wireless devices including IEEE 802.1 and IEEE 802.15.4 operate at ISM band. 

The current IEEE 802.15.6 standard defines three PHY layers, i.e., Narrowband (NB), Ultra wideband (UWB), and Human Body Communications (HBC) layers. The selection of each PHY depends on the application requirements. On the top of it, the standard defines a sophisticated MAC protocol that controls access to the channel. For time referenced resource allocations, the hub (or the coordinator) divides the time axis (or the channel) into a series of superframes. The superframes are bounded by beacon periods of equal length. To ensure high level security \cite{shahnaz}, the standard defines three levels: 1) level 0 - unsecured communication, 2) level 1 - authentication only, 3) level 2 - both authentication and encryption. In this paper, we briefly overview the PHY and MAC layers specifications together with the bandwidth efficiency of IEEE 802.15.6 standard for different frequency bands and data rates. We also discuss the security paradigm of the standard. 

The rest of the paper is organized into three sections. Section I discusses WBAN applications targeted by the standard. Section II presents the PHY and MAC layers specifications. This section also presents the bandwidth efficiency and the security paradigm of the standard. The final section concludes our work with useful remarks.
\begin{table*}[!t]
\renewcommand{\arraystretch}{1.3}
\caption{Modulation parameters for PLCP Header and PSDU}
\label{tab:1}
\centering
\begin{tabular}{|c|c|c|p{1.2cm}|p{1.2cm}|p{1.2cm}|}
\cline{1-6}
Frequency Band & Packet Component & Modulation & Symbol Rate (Kbps) & Code Rate BCH (n,k) & Information Data Rate (Kbps) \\ \cline{1-6}

\multicolumn{1}{|p{2cm}|}{402 - 405 MHz} & PLCP Header & $\pi$/2-DBPSK & 187.5 & (31,19) & 57.5     \\ \cline{2-6}
\multicolumn{1}{|p{2cm}|}{} & PSDU & $\pi$/2-DBPSK & 187.5 & (63,51) & 75.9     \\ \cline{2-6}
\multicolumn{1}{|p{2cm}|}{} & PSDU & $\pi$/4-DQPSK & 187.5 & (63,51) & 303.6     \\ \cline{1-6}

\multicolumn{1}{|p{2cm}|}{420 - 450 MHz} & PLCP Header & GMSK & 187.5 & (31,19) & 57.5     \\ \cline{2-6}
\multicolumn{1}{|p{2cm}|}{} & PSDU & GMSK & 187.5 & (63,51) & 75.9    \\ \cline{2-6}
\multicolumn{1}{|p{2cm}|}{} & PSDU & GMSK & 187.5 & (63,51) & 151.8    \\ \cline{1-6}

\multicolumn{1}{|p{2cm}|}{863 - 870 MHz} & PLCP Header & $\pi$/2-DBPSK & 250 & (31,19) & 76.6     \\ \cline{2-6}
\multicolumn{1}{|p{2cm}|}{} & PSDU & $\pi$/2-DBPSK & 250 & (63,51) & 101.2    \\ \cline{2-6}
\multicolumn{1}{|p{2cm}|}{} & PSDU & $\pi$/4-DQPSK & 250 & (63,51) & 404.8    \\ \cline{1-6}

\multicolumn{1}{|p{2cm}|}{902 - 928 MHz} & PLCP Header & $\pi$/2-DBPSK & 300 & (31,19) & 91.9     \\ \cline{2-6} 
\multicolumn{1}{|p{2cm}|}{} & PSDU & $\pi$/2-DBPSK & 300 & (63,51) & 121.4     \\ \cline{2-6} 
\multicolumn{1}{|p{2cm}|}{} & PSDU & $\pi$/4-DQPSK & 300 & (63,51) & 485.7    \\ \cline{1-6} 

\multicolumn{1}{|p{2cm}|}{950 - 956 MHz} & PLCP Header & $\pi$/2-DBPSK & 250 & (31,19) & 76.6      \\ \cline{2-6} 
\multicolumn{1}{|p{2cm}|}{} & PSDU & $\pi$/2-DBPSK & 250 & (63,51) & 101.2      \\ \cline{2-6} 
\multicolumn{1}{|p{2cm}|}{} & PSDU & $\pi$/4-DQPSK & 250 & (63,51) & 404.8      \\ \cline{1-6}

\multicolumn{1}{|p{2cm}|}{2360-2400 MHz} & PLCP Header & $\pi$/2-DBPSK & 600 & (31,19) & 91.9      \\ \cline{2-6} 
\multicolumn{1}{|p{2.2cm}|}{2400-2483.5 MHz} & PSDU & $\pi$/2-DBPSK & 600 & (63,51) & 121.4     \\ \cline{2-6} 
\multicolumn{1}{|p{2cm}|}{} & PSDU & $\pi$/2-DBPSK & 600 & (63,51) & 485.7      \\ \cline{1-6}

\end{tabular}
\end{table*}
\section{Target Applications}
\begin{figure}[!h]
\centering
\includegraphics[width=3in]{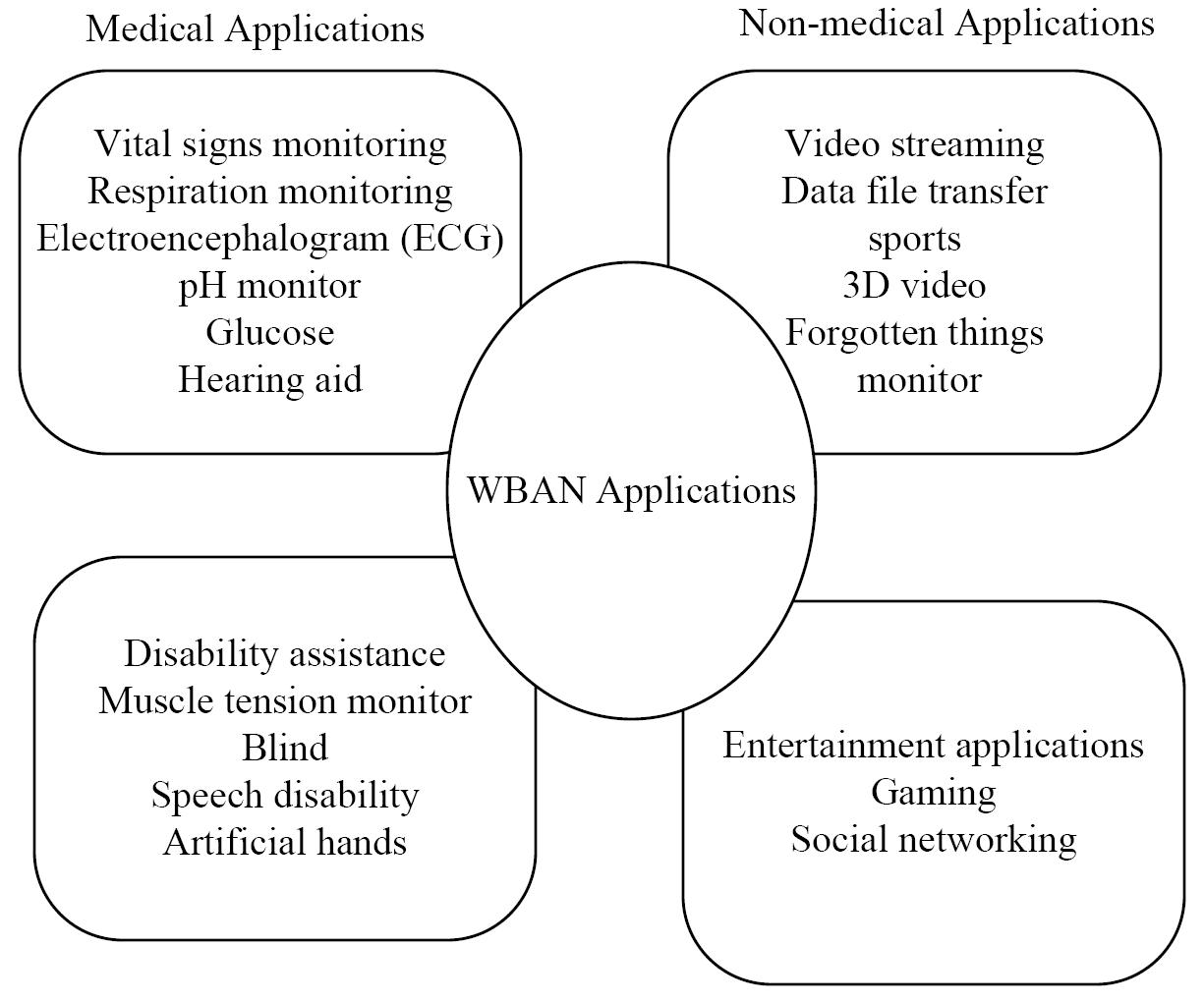}
\caption{WBAN applications}
\label{fig:2}
\end{figure}

The WBAN applications targeted by the IEEE 802.15.6 standard are divided into medical and non-medical applications as given in Fig. \ref{fig:2}. Medical applications include collecting vital information of a patient continuously and forward it to a remote monitoring station for further analysis. This huge amount of data can be used to prevent the occurrence of myocardial infarction and treat various diseases such as gastrointestinal tract, cancer, asthma, and neurological disorder. WBAN can also be used to help people with disabilities. For example, retina prosthesis chips can be implanted in the human eye to see at an adequate level. Non-medical applications include monitoring forgotten things, data file transfer, gaming, and social networking applications. In gaming, sensors in WBAN can collect coordinates movements of different parts of the body and subsequently make the movement of a character in the game, e.g., moving soccer player or capturing the intensity of a ball in table tennis. The use of WBAN in social networking allows people to exchange digital profile or business card only by shaking hands.         

\begin{figure*}[!t]
\centering
\includegraphics[width=5in]{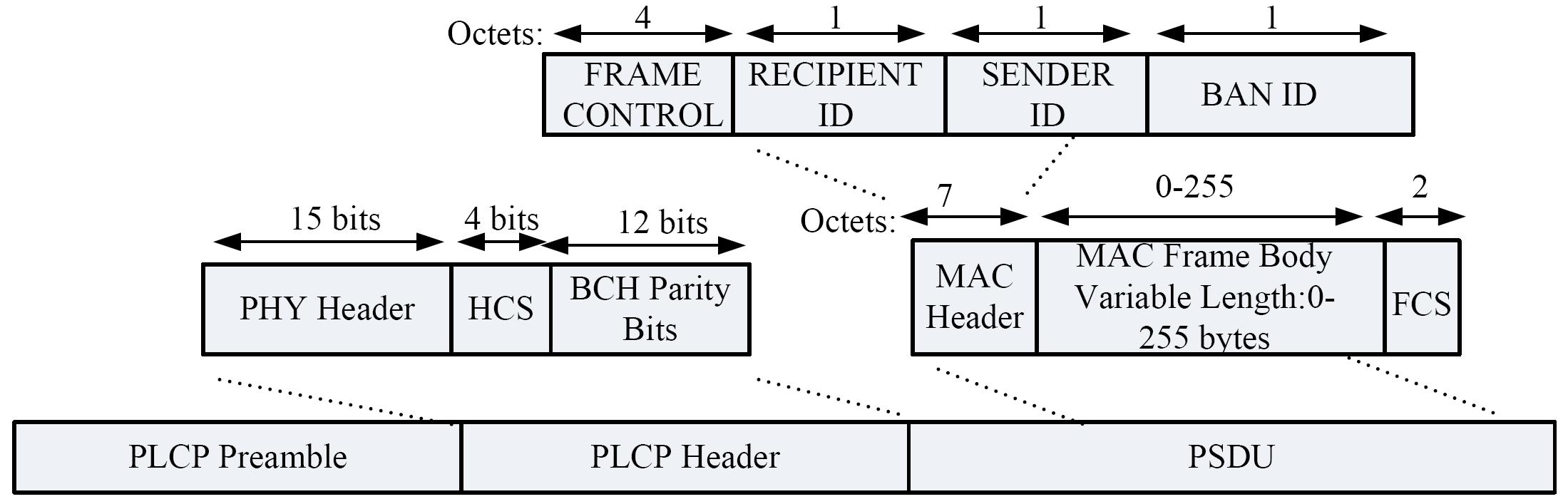}
\caption{IEEE 802.15.6 NB PPDU structure}
\label{fig:3}
\end{figure*}
\begin{figure*}[!t]
\centering
\includegraphics[width=4in]{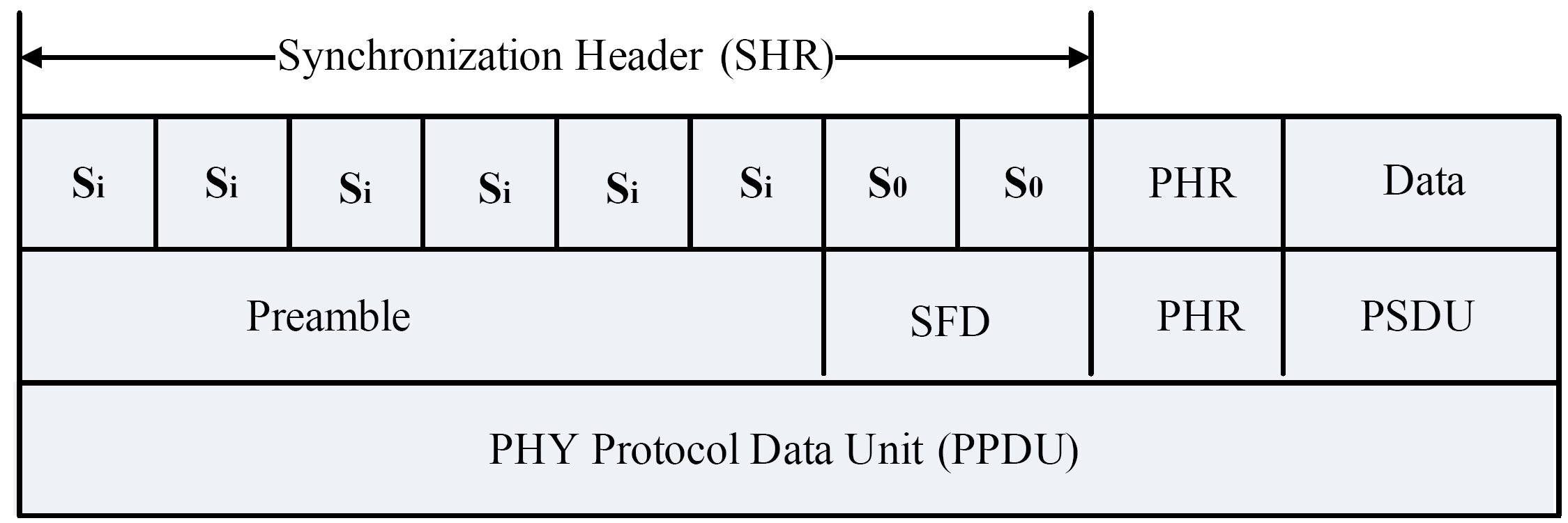}
\caption{IEEE 802.15.6 UWB PPDU structure}
\label{fig:4}
\end{figure*}
\begin{figure*}[!t]
\centering
\includegraphics[width=4in]{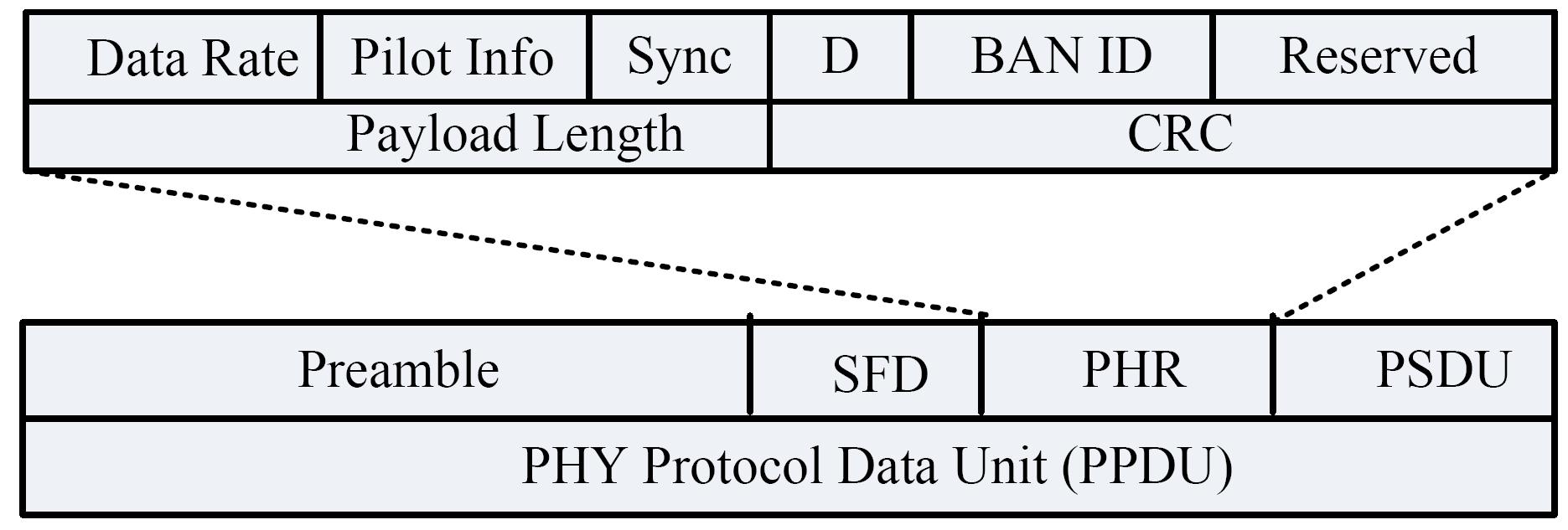}
\caption{IEEE 802.15.6 EFC PPDU structure}
\label{fig:5}
\end{figure*}\section{IEEE 802.15.6 Description}
The following sections describe the main features of IEEE 802.15.6 standard including PHY and MAC layers specifications.
\subsection{PHY Layer Specification}
As mentioned earlier, the IEEE 802.15.6 supports three different PHYs, i.e., NB, UWB, and HBC.
\subsubsection{Narrowband PHY (NB)}
The NB PHY is responsible for activation/deactivation of the radio transceiver, Clear Channel Assessment (CCA) within the current channel and data transmission/reception. The Physical Protocol Data Unit (PPDU) frame of NB PHY contains a Physical Layer Convergence Procedure (PLCP) preamble, a PLCP header, and a PHY Service Data Unit (PSDU) as given in Fig \ref{fig:3}. The PLCP preamble helps the receiver in the timing synchronization and carrier-offset recovery. It is the first component being transmitted. The PLCP header conveys information necessary for a successful decoding of a packet to the receiver. The PLCP header is transmitted after PLCP preamble using the given header data rate in the operating frequency band. The last component of PPDU is PSDU which consists of a MAC header, MAC frame body, Frame Check Sequence (FCS) and is transmitted after PLCP header using any of the available data rates in the operating frequency band. A WBAN device should be able to support transmission and reception in one of the frequency bands summarized in Table \ref{tab:1}. The table further shows the data-rate dependent modulations parameters for PLCP header and PSDU. In NB PHY, the standard uses Differential Binary Phase-shift Keying (DBPSK), Differential Quadrature Phase-shift Keying (DQPSK), and Differential 8-Phase-shift Keying (D8PSK) modulation techniques except
420-450 MHz which uses a Gaussian minimum shift keying (GMSK) technique.  
\subsubsection{Ultra Wideband  PHY (UWB)}
UWB PHY operates in two frequency bands: low band and high band. Each band is divided into channels, all of them characterized by a bandwidth of 499.2 MHz. The low band consists of 3 channels (1-3) only. The channel 2 has a central frequency of 3993.6 MHz and is considered a mandatory channel. The high band consists of eight channels (4-11) where channel 7 with a central frequency 7987.2 MHz  is considered a mandatory channel, while all other channels are optional. A typical UWB device should support at least one of the mandatory channels. The UWB PHY transceivers allow low implementation complexity and generate signal power levels in the order of those used in the MICS band.

Fig. \ref{fig:4} shows the UWB PPDU that contains a Synchronization Header (SHR), a PHY Header (PHR), and PSDU. The SHR is composed of a preamble and a Start Frame Delimiter (SFD). The PHR conveys information about the data rate of the PSDU, length of the payload and scrambler seed. The information in the PHR is used by the receiver in order to decode the PSDU. The SHR is formed of repetitions of Kasami sequences of length 63. Typical data rates range from 0.5 Mbps up to 10 Mbps with 0.4882 Mbps as the mandatory one.
\subsubsection{Human Body Communications PHY (HBC)}
HBC PHY operates in two frequency bands centered at 16 MHz and 27 MHz with the bandwidth of 4 MHz. Both operating bands are valid for the United States, Japan, and Korea, and the operating band at 27MHz is valid for Europe. HBC is the Electrostatic Field Communication (EFC) specification of PHY, which covers the entire protocol for WBAN such as packet structure, modulation, preamble/SFD, etc.  Fig. \ref{fig:5} describes the PPDU structure of EFC that is composed of a preamble, SFD, PHY header and PSDU. The preamble and SFD are fixed data patterns. They are pre-generated and sent ahead of the packet header and payload. The preamble sequence is transmitted four times in order to ensure packet synchronization while the SFD is transmitted only once. When the packet is received by the receiver, it finds the start of the packet by detecting the preamble sequence, and then it finds the start of the frame by detecting the SFD.
\begin{figure*}[!t]
\centering
\includegraphics[width=5in]{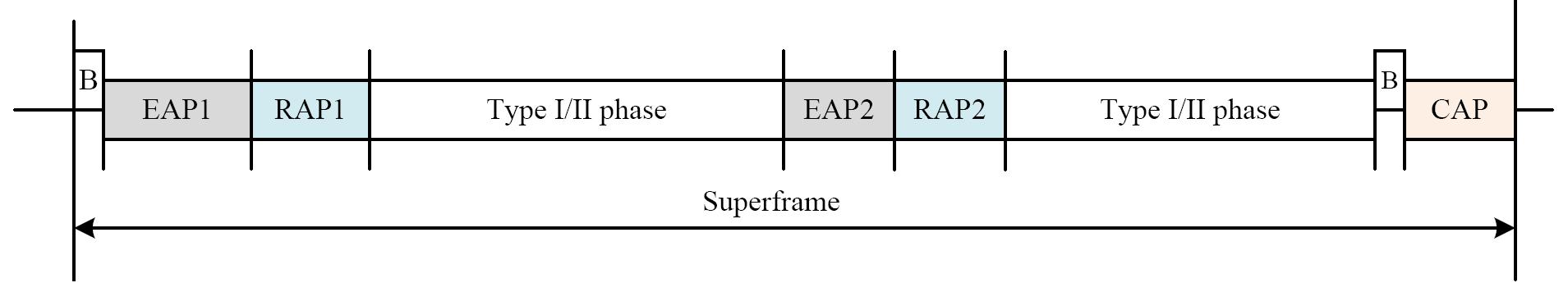}
\caption{IEEE 802.15.6 superframe structure}
\label{fig:6}
\end{figure*}
\begin{figure*}[!t]
\centering
\includegraphics[width=5.5in]{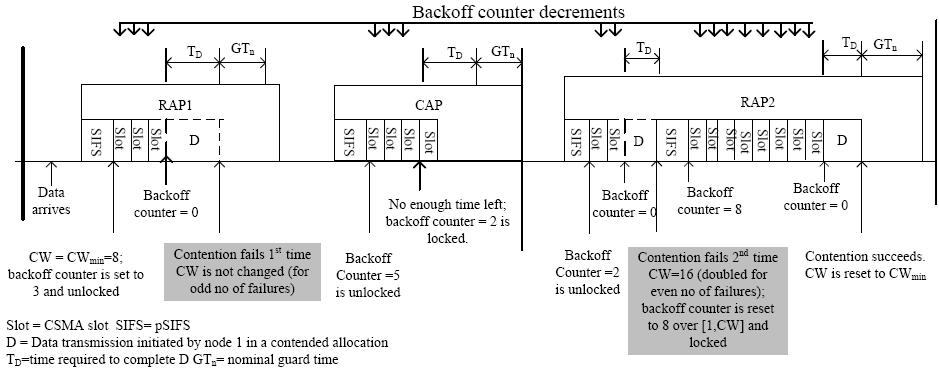}
\caption{CSMA/CA procedure in IEEE 802.15.6 standard}
\label{fig:7}
\end{figure*}
\subsection{MAC Layer Specification}
In IEEE 802.15.6, the entire channel is divided into superframe structures. Each superframe is bounded by a beacon period of equal length. The hub selects the boundaries of the beacon period and thereby selects the allocation slots. The hub may also shift the offsets of the beacon period. Generally, the beacons are transmitted in each beacon period except in inactive superframes or unless prohibited by regulations such as in MICS band. The IEEE 802.15.6 network operates in one of the following modes.

\subsubsection{Beacon mode with beacon period superframe boundaries}
In this mode, the beacons are transmitted by the hub in each beacon period except in inactive superframes or unless prohibited by regulations. Fig. \ref{fig:6} shows the superframe strucutre of IEEE 802.15.6, which is divided into Exclusive Access Phase 1 (EAP1), Random Access Phase 1 (RAP1), Type I/II phase, Exclusive Access Phase 2 (EAP 2), Random Access Phase 2 (RAP 2), Type I/II phase, and a Contention Access Phase (CAP). In EAP, RAP and CAP periods, nodes contend for the resource allocation using either CSMA/CA or a slotted Aloha access procedure.  The EAP1 and EAP2 are used for highest priority traffic such as reporting emergency events. The RAP1, RAP2, and CAP are used for regular traffic only. The Type I/II phases are used for uplink allocation intervals, downlink allocation intervals, bilink allocation intervals, and delay bilink allocation intervals. In Type I/II phases, polling is used for resource allocation. Depending on the application requirements, the coordinator can disable any of these periods by setting the duration length to zero. 

\subsubsection{Non-beacon mode with superframe boundaries}
In this mode, the entire superframe duration is covered either by a type I or a type II access phase but not by both phases.
\begin{figure*}[!t]
\centering
\includegraphics[width=3.5in]{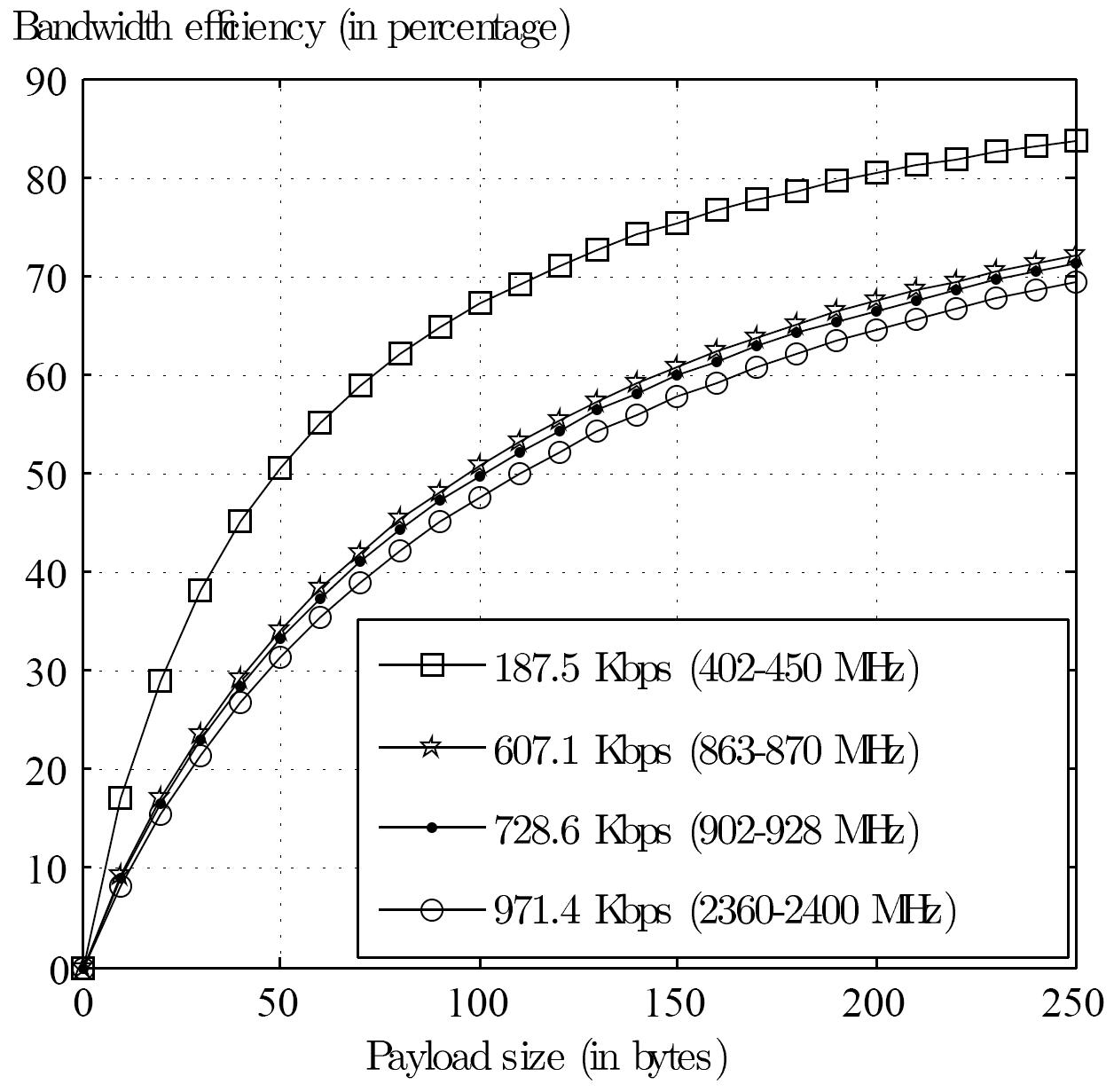}
\caption{Bandwidth efficiency of IEEE 802.15.6}
\label{fig:8}
\end{figure*}
\begin{figure*}[!t]
\centering
\includegraphics[width=5in]{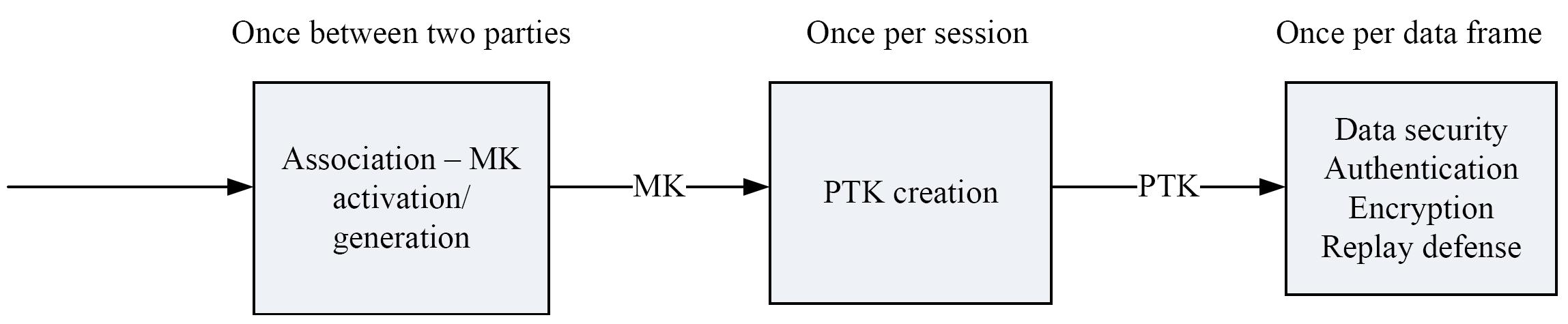}
\caption{Security structure of IEEE 802.15.6}
\label{fig:9}
\end{figure*}
\subsubsection{Non-beacon mode without superframe boundaries}
In this mode, the coordinator provides unscheduled Type II polled allocation only.

The access mechanisms used in each period of the superframe are divided into three categories: 1) Random access mechanism, which uses either CSMA/CA or a slotted Aloha procedure for resource allocation, 2) Improvised and unscheduled access (connectionless contention-free access), which uses unscheduled polling/posting for resource allocation, and 3) Scheduled access and variants (connection-oriented contention-free access),  which schedules the allocation of slots in one or multiple upcoming superframes, also called 1-periodic or m-periodic allocations. These mechanisms are comprehensively discussed in the standard. Here we explain the basic procedures of the CSMA/CA protocol defined in the standard. In CSMA/CA, a node sets its backoff counter to a random integer number uniformly distributed over the interval $[1, CW]$ where $CW \in (CW_{min}, CW_{max})$. The values of of $CW_{min}$ and $CW_{max}$ vary depending on the user priorities. The node starts decrementing the backoff counter by one for each idle CSMA slot of duration equal to $pCSMASlotLength$. The data is transmitted when the backoff counter reaches zero. If the channel is busy because of a frame transmission, the node locks its backoff counter until the channel is idle. The $CW$ is doubled for even number of failures (when the node fails to receive an acknowledgement or group
acknowledgement) until it reaches $CW_{max}$. Fig. \ref{fig:7} shows the CSMA/CA procedure defined in the IEEE 802.15.6 standard. In RAP1, the node first waits for $SIFS=pSIFS$ duration and then unlocks the backoff counter until it reaches zero where the data transmission starts. But the node fails to receive an acknowledgement and the contention fails. According to the standard, the $CW$ is not doubled for odd number of failures and therefore it remains unchanged. In CAP, the node sets the backoff counter to 5 and locks it at 2 since the time between the end of the slot and the end of the CAP is not enough for completing the data transmission and the Nominal Guard Time, represented by $GT_{n}$. The backoff counter is unlocked in the RAP2 period. Again the node fails to receive an acknowledgement and the contention fails. The $CW$ gets doubled (for even number of failures) and the backoff counter is set to 8. Once the data transmission is successful, the $CW$ is set to $CW_{min}$. Further details about the CSMA/CA procedure can be found in the standard \cite{802.15.6}.

To measure the spectral utilization of IEEE 802.15.6 when using CSMA/CA procedure, we analyze the bandwidth efficiency for different frequency bands and data rates. The bandwidth efficiency is inversely proportional to the basic data rate. The following assumptions are considered during the calculations. 1) Bit Error Rate (BER) is zero, 2) The channel is perfect with no losses due to collisions, 3) The node always has a packet to send, 4) There are no packet losses due to buffer overflow. Fig. \ref{fig:8} presents the bandwidth efficiency as a function of payload size. It can be seen that the efficiency increases as we increase the payload size, i.e., it is 83.6\% for 187.5 Kbps and 69.4\% for 971 Kbps, respectively. These results can help the application protocol designer to see the effects of the payload size on the bandwidth efficiency and can be used to minimize jitter in multimedia applications. Detailed analysis can be found in \cite{sana-wcnc2011}.

\subsection{Security Paradigm}
The standard defines the following three levels of security. Each security level has different security properties, protection levels and frame formats.
\subsubsection{Level 0 - unsecured communication}
This is the lowest security level where data is transmitted in unsecured frames. There is no mechanism for data authentication and integrity, confidentiality and privacy protection, and replay defense.
\subsubsection{Level 1 - authentication only}
This is the medium security level where data is transmitted in secured authentication but is not encrypted. The confidentially and privacy is not supported by this mode.
\subsubsection{Level 3 - authentication and encryption}
This is the highest security level where data is transmitted in secured authentication and encryption frames. It provides solutions to all of the problems not covered by the level 0 and level 1. 

The required security level is selected during the association process, i.e., when a node is joining the network. For unicast communication, a pre-shared Master Key (MK) or a new key (established via unauthenticated association) is activated. Then a Pairwise Temporal Key (PTK) is established, which is used once per session. For multicast communication, a Group Temporal Key (GTK) is shared with the corresponding multicast group. The whole security structure is given in Fig. \ref{fig:9}.   

\section{Conclusion}
This paper presented a brief overview of the new IEEE 802.15.6 standard. We studied the PHY and MAC layers specifications and identified their key points. In addition, we analyzed the bandwidth efficiency of the standard for CSMA/CA procedure. The efficiency results were presented for different frequency bands and data rates. We observed that increase in the payload size improves the bandwidth efficiency. We also highlighted different security modes of the standard. This paper can be used to quickly understand the key concepts of IEEE 802.15.6 without reading the whole standard.   


\section*{Acknowledgment}
This work was supported by the National Research Foundation of Korea (NRF) grant funded by the Korea government (MEST)(No. No.2010-0018116). 


%

\end{document}